\documentclass[aps,twocolumn,showkeys,showpacs]{revtex4}
\usepackage{amsfonts}
\usepackage{amsmath}
\usepackage{amssymb}
\usepackage{graphicx}
\usepackage{hyperref}

\begin{document}

\title{USING ENERGY CONDITIONS TO DISTINGUISH BRANE MODELS AND STUDY BRANE MATTER}
\author{Yongli Ping\footnote{ylping@student.dlut.edu.cn}, Lixin Xu\footnote{
lxxu@dlut.edu.cn}, Baorong Chang and Hongya Liu} \affiliation{School
of Physics and Optoelectronic Technology, Dalian University of
Technology, Dalian, Liaoning 116024, P.R.China} 

\begin{abstract}
Current universe (assumed here to be normal matter on the brane) is
pressureless from observations. In this case the energy condition is
$\rho_0\geq0$ and $p_0=0$. By using this condition, brane models can
be distinguished. Then, assuming arbitrary component of matter in
DGP model, we use four known energy conditions to study the matter
on the brane. If there is nonnormal matter or energy (for example
dark energy with $w<-1/3$) on the brane, the universe is
accelerated.
\end{abstract}
\maketitle
\section{Introduction}
In braneworld scenarios, our universe is a $3$-brane embedded in a
higher-dimensional bulk \cite{Arkani}-\cite{Randall}. It is proposed
that braneworld modification of gravity to explain the accelerating
expansion of the universe by Dvali, Gabadadze and Porrati (DGP)
\cite{DGP}. (see \cite{Lue} for a recent review). In this model our
universe is a 3-brane embedded in an infinite-volume extra space.
Meanwhile, a more general class of braneworld models is described by
Sahni and Shtanov \cite{Sahni}. The cosmological constant in the
bulk and the curvature term in the action for the brane with
coefficient $m^2$ are included in this model. Then the exact global
solutions of brane universes are given in \cite{Liu}. It contains
two arbitrary functions of time $t$. As we know, a number of recent
observations indicate that the expansion rate of our universe is
accelerating \cite{Riess}. The cosmological solution of DGP theory
exhibits self-acceleration on the brane \cite{Deffayet}. This
solution describes a universe that is accelerating beyond the
crossover scale. Moreover, the braneworld models of dark energy are
studied in \cite{Sahni2}. Now the acceleration behavior in
braneworld scenarios are widely studied in terms of current
observational data \cite{Song}.

The purpose of this paper is to use energy conditions to distinguish
the different brane models and study the matter in DGP brane. As we
know in the braneworld model, the standard matter particles and
forces are confined on the 3-brane, while gravity can freely
propagate in all dimensions. From current observation, we can
distinguish these brane models by energy conditions. Then, in DGP
brane we assume there is normal matter or nonnormal and apply energy
conditions to study the matter on the brane.

\section{Using energy conditions to distinguish the brane universes}

In braneworld model, the standard matter particles and forces are
confined on the 3-brane while gravity can freely propagate in all
dimensions. The Hubble and deceleration parameters are respectively
given on $y=0$ brane by
\begin{equation}
H=\frac{\dot{a}}{a},\ \ \
q=-\frac{1}{H^2}\frac{\ddot{a}}{a},\label{q1}
\end{equation}
 The conservation
equation for the energy-momentum tensor of the cosmic fluid is
\begin{equation}
\dot{\rho}+3H(p+\rho)=0.\label{ce}
\end{equation}

We note that observations show that current universe (assumed here
to be normal matter, as opposed to dark energy) is pressureless. In
this case the energy condition is $\rho_0\geq0$ and $p_0=0$, where
the subindex $0$ means the current quantity. For different brane
models, we can obtain different constraints with this energy
condition. The analysis is given as follows.

\subsection{DGP brane}
For the DGP brane model,\cite{DGP} the Friedmann equation is written
as \cite{Deffayet}
\begin{equation}
H^2+\frac{k}{a^2}=(\sqrt{\rho/{3M^2_{pl}}+1/{4r^2_c}}+\epsilon\frac{1}{2r_c})^2,\label{H-DGP}
\end{equation}
where $\rho$ is the total cosmic fluid energy density, and
$\epsilon=\pm1$. $r_c=M_{pl}^2/{2M_{(5)}^3}$, $M_{pl}$ and $M_{(5)}$
are independent parameters (in general there could be an relation
between the two quantity ). For the self-inflationary solution, it
was adopted $\epsilon=1$. So we set $\epsilon=1$ firstly. For $k=0$,
Eq. (\ref{H-DGP}) is rewritten as
\begin{equation}
\frac{\rho}{3M^2_{pl}}=H^2-\frac{H}{r_c},\label{rho1}
\end{equation}
with $H\geq1/{2r_c}$. Then from $\rho_0\geq0$, the inequality is
obtained as $ H_0\geq{1}/{r_c}$. This inequality gives a relation
between $H_0$ and $r_c$. Since $H$ is observable quantity and $r_c$
is a quantity from the theory model, this inequality is beneficial
to choose $r_c$. From the conservation equation (\ref{ce}) and the
Friedmann equation (\ref{rho1}), the pressure is expressed as
\begin{equation}
\frac{p}{M^2_{pl}}=H^2(2q-1)-\frac{H}{r_c}(q-2).\label{p1}
\end{equation}
By using $p_0=0$, we get the relation
\begin{equation}
\frac{1}{r_c}=\frac{H_0(2q_0-1)}{q_0-2}.\label{r1}
\end{equation}
From the inequality $ H_0\geq{1}/{r_c}>0$ and the Eq. (\ref{r1}), we
have $-1\leq{q_0}<1/2$. According current observation value
$q_0=-0.81\pm0.14$ \cite{Rapetti}, this inequality is valid.

Then, if we choose $\epsilon=-1$, with $H\geq-1/{2r_c}$ the density
and pressure are
\begin{eqnarray}
\frac{\rho}{3M^2_{pl}}&=&H^2+\frac{H}{r_c},\\
\frac{p}{M^2_{pl}}&=&H^2(2q-1)+\frac{H}{r_c}(q-2).
\end{eqnarray}
Since the energy conditions $\rho_0\geq0$ and $p_0=0$, we have two
relations as
\begin{eqnarray}
&&H_0+\frac{1}{r_c}\geq0,\label{H02}\\
&&H_0(2q_0-1)+\frac{1}{r_c}(q_0-2)=0.\label{r02}
\end{eqnarray}
So we can get $-1\leq{q}_0<2$. Since $H_0>0$, from (\ref{r02}), it
is obtained that $1/2<q_0<2$. This deceleration parameter shows our
universe should be decelerating expansion. However according to the
present observation this is not true, and our universe is
accelerating. Therefore, to satisfy the energy conditions,
$\epsilon\neq-1$ in this model.

\subsection{Brane models with brane tension and cosmological constant}
For the braneworld models in \cite{Shtanov}, the Friedmann equation
takes the form
\begin{equation}
H^2+\frac{k}{a^2}=\frac{\rho+\sigma}{3m^2}+
\frac{2}{l^2}\left[1\pm\sqrt{1+l^2(\frac{\rho+\sigma}{3m^2}-\frac{\Lambda}{6}-\frac{C}{a^4})}\right],\label{H}
\end{equation}
where $\sigma$ is the brane tension, $\Lambda$ is the bulk
cosmological constant, $l=m^2/M^3$, $C$ is a constant, and the term
$C/a^4$ plays the role of ``dark radiation''. Next, we set $k=0$,
$\Lambda=0$ and neglect the radiation density. Eq. (\ref{H}) is
rewritten as
\begin{equation}
H^2=\frac{\rho+\sigma}{3m^2}+
\frac{2}{l^2}\left[1\pm\sqrt{1+l^2\frac{\rho+\sigma}{3m^2}}\right].\label{Hr}
\end{equation}
In this model, the Friedmann equation contains $\pm$ sign, so this
model is separated two types, i.e. BRANE1 and BRANE2 with ``$-$" and
``$+$" respectively. Next, we will discuss them as follows:

For BRANE1, choosing the ``$-$" sign, the Friedmann equation is
\begin{equation}
H^2=\frac{\rho+\sigma}{3m^2}+
\frac{2}{l^2}\left[1-\sqrt{1+l^2\frac{\rho+\sigma}{3m^2}}\right].\label{Hr}
\end{equation}
When the condition is
\begin{equation}
1-(H^2-\frac{\rho+\sigma}{3m^2})\frac{l^2}{2}\geq0,\label{c21}
\end{equation}
the density is
\begin{equation}
\rho=3m^2(H^2\pm\frac{2}{l}H)-\sigma,\label{rho21}
\end{equation}
where, if $H^2-(\rho+\sigma)/{3m^2}\geq0$, adopt ``$-$"; while if
$H^2-(\rho+\sigma)/{3m^2}\leq0$, adopt ``$+$". These can ensure
mathematical reasonability of our calculation.

Firstly, choosing ``$-$" sign, the density and pressure are
\begin{eqnarray}
\rho&=&3m^2(H^2-\frac{2}{l}H)-\sigma,\label{rho211}\\
p&=&m^2[H^2(2q-1)-\frac{2}{l}H(q-2)]+\sigma.\label{p211}
\end{eqnarray}
Because the inequality $\rho_0\geq0$ and the equality $p_0=0$, we
get
\begin{eqnarray}
&&3m^2(H_0^2-\frac{2}{l}H_0)-\sigma\geq0,\label{ie21}\\
&&m^2[H_0^2(2q_0-1)-\frac{2}{l}H_0(q_0-2)]+\sigma=0.\label{e21}
\end{eqnarray}
Substituted Eq. (\ref{e21}) into the inequality (\ref{ie21}), one
new inequality can be obtained as
\begin{equation}
(H_0-\frac{1}{l})(q_0+1)\geq0.
\end{equation}
Therefore, we get two solutions of this inequality: one is
$H_0\geq1/l$ and $q_0\geq-1$, the other is $H_0\leq1/l$ and
$q_0\leq-1$. In terms of current observation, $q_0=-0.81\pm0.14$
therefor we find $H_0\geq1/l$ and $q_0\geq-1$. However, the
inequality (\ref{c21}) and equation (\ref{rho211}) imply $H\leq1/l$.
So, only the critical situation $H_0=1/l$ is suitable, but this is a
critical condition.

Secondly, choosing ``$+$" sign, we obtain the density and pressure
as
\begin{eqnarray}
\rho&=&3m^2(H^2+\frac{2}{l}H)-\sigma,\label{ie22}\\
p&=&m^2[H^2(2q-1)+\frac{2}{l}H(q-2)]+\sigma.\label{e22}
\end{eqnarray}
Form the energy conditions $\rho_0\geq0$ and $p_0=0$, they are
obtained as
\begin{eqnarray}
&&3m^2(H_0^2+\frac{2}{l}H_0)-\sigma\geq0,\label{ie212}\\
&&m^2[H_0^2(2q_0-1)+\frac{2}{l}H_0(q_0-2)]+\sigma=0.\label{e212}
\end{eqnarray}
Substituting Eq. (\ref{e212}) into the inequality (\ref{ie212}), we
have
\begin{equation}
(H_0+\frac{1}{l})(q_0+1)\geq0.
\end{equation}
Since $H_0>0$ and $l>0$, we get the solution of this inequality as
$q_0\geq-1$. In terms of current observation, $q_0=-0.81\pm0.14$,
therefore $q_0\geq-1$ is valid. At the same time, from (\ref{e212}),
we can find $\sigma\geq0$. This is a positive brane tension.

For BRANE2, taking ``$+$" sign, the Friedmann equation is
\begin{equation}
H^2=\frac{\rho+\sigma}{3m^2}+
\frac{2}{l^2}\left[1+\sqrt{1+l^2\frac{\rho+\sigma}{3m^2}}\right].\label{Hr}
\end{equation}
When the condition is
\begin{equation}
(H^2-\frac{\rho+\sigma}{3m^2})\frac{l^2}{2}-1\geq0,\label{c22}
\end{equation}
the density is written as
\begin{equation}
\rho=3m^2(H^2\pm\frac{2H}{l})-\sigma
\end{equation}
where, if $H^2-(\rho+\sigma)/{3m^2}\geq0$, we will take ``$-$";
while if $H^2-(\rho+\sigma)/{3m^2}\leq0$, we will take ``$+$" for
being reasonable in mathematics.

Firstly, taking ``$-$" sign, we have
\begin{eqnarray}
\rho&=&3m^2(H^2-\frac{2}{l}H)-\sigma.\label{rho221}\\
p&=&m^2[H^2(2q-1)-\frac{2}{l}H(q-2)]+\sigma.\label{p221}
\end{eqnarray}
Eq.(\ref{rho221}) and Eq.(\ref{c22}) imply the condition
$Hl-1\geq0$. Under the energy conditions $\rho_0\geq0$ and $p_0=0$,
they are obtained as
\begin{eqnarray}
&&3m^2(H_0^2-\frac{2}{l}H_0)-\sigma\geq0.\label{ie221}\\
&&m^2[H_0^2(2q_0-1)-\frac{2}{l}H_0(q_0-2)]+\sigma=0.\label{e221}
\end{eqnarray}
Substituting Eq. (\ref{e221}) into the inequality (\ref{ie221}), we
have
\begin{equation}
(H_0-\frac{1}{l})(q_0+1)\geq0.
\end{equation}
Then, we get two solutions: one is $H_0\geq{1}/{l}$ when
$q_0\geq-1$; the other is $H_0\leq{1}/{l}$ when $q_0\leq-1$. From
the current value $q_0=-0.81\pm0.14$, the latter solution is not
satisfied. So, the suitable solution is $H_0\geq{1}/{l}$ when
$q_0\geq-1$. However in this situation, from (\ref{e221}), since
$\sigma$ is determined by $l$, we can not verify it is positive or
negative.

Secondly, taking ``$+$" sign, the density and the pressure are
\begin{eqnarray}
\rho&=&3m^2(H^2+\frac{2}{l}H)-\sigma,\label{rho222}\\
p&=&m^2[H^2(2q-1)+\frac{2}{l}H(q-2)]+\sigma.\label{p222}
\end{eqnarray}
Considering the energy conditions $\rho_0\geq0$ and $p_0=0$, we
obtain
\begin{eqnarray}
&&3m^2(H_0^2+\frac{2}{l}H_0)-\sigma\geq0.\label{ie222}\\
&&m^2[H_0^2(2q_0-1)+\frac{2}{l}H_0(q_0-2)]+\sigma=0.\label{e222}
\end{eqnarray}
Substituting Eq. (\ref{e222}) into the inequality (\ref{ie222}), the
new inequality is
\begin{equation}
(H_0+\frac{1}{l})(q_0+1)\geq0.
\end{equation}
for $H_0>0$ and $l>0$, the solution of this inequality is
$q_0\geq-1$. From the current value $q_0=-0.81\pm0.14$, this
satisfies the $q_0\geq-1$. But from Eq.(\ref{rho222}) and
(\ref{c22}), it is obtained that $-H_0l_0-1\geq0$. This is
incompatible with $H_0>0$. Therefore there is no valid solution.

\section{Study matter on DGP brane with energy conditions}
In this Section I, $\epsilon=1$ DGP brane is considered. Now we
assume arbitrary matter besides normal matter on the brane. From
(\ref{rho1}) and (\ref{p1}), the $\rho$ and $p$ with $M^2_{pl}=1$
are described as
\begin{eqnarray}
\rho&=&3(H^2-\frac{H}{r_c})\label{1rho}\\
p&=&H^2(2q-1)-\frac{H}{r_c}(q-2)\label{1p}.
\end{eqnarray}

The standard classical energy conditions are the null energy
condition (NEC), weak energy condition (WEC), strong energy
condition (SEC), and dominant energy condition (DEC). Basic
definitions of these energy conditions can be found in Ref.
\cite{Wald}. For the case in cosmology they are
\begin{eqnarray}
NEC &:&\qquad \rho +p\geq 0,  \label{nec} \\
WEC &:&\qquad \rho \geq 0\qquad\mbox{and}\quad \rho +p\geq 0,  \label{wec} \\
SEC &:&\qquad \rho +3p\geq 0\qquad\mbox{and}\quad \rho +p\geq 0,
\label{sec}\\
DEC &:&\qquad \rho \geq 0\qquad\mbox{and}\quad \rho \geq \left\vert
p\right\vert .  \label{dec}
\end{eqnarray}
They were used in deriving many theorems such as the singularity
theorems \cite{Howking}, the censorship theorem \cite{Censorship}
and so on. Other applications of the energy conditions to cosmology
can be found in \cite{Visser97}-\cite{Bergliaffa}. In the classical
general reletivity these energy conditions are satisfied, while if
considering the quantum effect, energy conditions should be
violated\cite{Hossain,Zhu}.

We consider the universe is dominated by one fluid with $\rho$ and
$p$. The equation of state is
\begin{equation}
w=\frac{p}{\rho},
\end{equation}
where $w$ may be arbitrary form on the brane. Substituting this
equation in to the four energy conditions, we get
\begin{eqnarray}
NEC &:&\qquad (1+w)\rho \geq 0,  \label{nec-w} \\
WEC &:&\qquad \rho \geq 0\quad \text{and}\quad w\geq -1,  \label{wec-w} \\
SEC &:&\qquad (1+3w)\rho \geq 0\quad \text{and}\quad (1+w)\rho \geq
0,
\label{sec-w} \\
DEC &:&\qquad \rho \geq 0\quad \text{and}\quad \left\vert
w\right\vert \leq 1.  \label{dec-w}
\end{eqnarray}
Following previous usage (see, for example \cite{Visser97}) we call
matter that satisfies all the four energy conditions ``normal'' and
call matter that specifically violates the SEC ``abnormal''. And we
call matter that violates any one of the four energy conditions
``non-normal''. In the following we will discuss the normal and
nonnormal matter respectively.

Firstly, for normal matter all the four standard energy conditons
should be satisfied. From (\ref{nec-w})-(\ref{dec-w}) we have
\begin{equation}
\text{Normal Matter}:\qquad \rho \geq 0\quad \text{and}\quad
-1/3\leq w\leq 1.  \label{NM}
\end{equation}
With the use of these constraints on the brane, we obtain from
(\ref{1rho}) and (\ref{1p}) that is
\begin{eqnarray}
H&\geq&\frac{1}{r_c}\label{Hnm},\\
q&\geq&-\frac{3}{r_c}\frac{1}{2H-\frac{1}{r_c}}\label{qnm-},\\
q&\leq&\frac{4H-\frac{5}{r_c}}{2H-\frac{1}{r_c}}\label{qnm}.
\end{eqnarray}
Therefore, when $1/r_c\leq{H}\leq5/(4r_c)$, the deceleration $q$
satisfies $-3\leq{q}\leq0$ and this show the universe is
accelerating. Even when $H\geq5/(4r_c)$, from (\ref{qnm-}), we can
not exclude that $q$ is negative.

Secondly, we discuss "nonnormal" matter on the brane. The
constraints are described as
\begin{equation}
\rho \geq 0\quad \text{and}\quad {w}\leq-\frac{1}{3}. \label{AN}
\end{equation}
Then, we have
\begin{eqnarray}
H&&\geq\frac{1}{r_c}\label{Hnnm},\\
{q}&&\leq-\frac{3}{r_c}\frac{1}{2H-\frac{1}{r_c}}\label{qnnm-}.
\end{eqnarray}
Since $r_c\geq0$ and ${2H-{1}/{r_c}}\geq0$, the deceleration $q$ is
negative. If there are ``abnormal" or ``nonnormal" matter on the
brane, the universe do be accelerating.

If normal matter in the standard FRW cosmology, all energy
conditions are satisfied and the universe is decelerating. However,
if normal matter on DGP brane, all energy conditions are also
satisfied but the accelerating universe can be obtained. Moreover,
the EOS of quintessence field is $-1\leq{w}\leq1$. Therefore
quintessence is a normal matter if $-1/3\leq{w}<1$ and abnormal
matter if $-1\leq{w}<-1/3$. If quintessence with $-1\leq{w}<-1/3$ on
the brane, the universe is accelerated. Because the EOS of phantom
is $w<-1$, the universe is accelerated by this field on the brane.

\section{Conclusion}
In this paper, energy conditions are used to distinguish different
brane models and study the matter on DGP brane. We notice that
observations show that the current of universe (assumed here to be
normal matter, as opposed to dark energy) is pressureless.
Therefore, the energy consitions reduce to the inequality
$\rho_0\geq0$ and the equation $p_0=0$. Then we use these two
relations to analyze DGP model and the models with brane tension and
cosmological constant. In the DGP model, we find when $\epsilon=1$,
the energy conditions are satisfied, while they does not satisfy the
$\epsilon=-1$ case. For the models with brane tension and
cosmological constant, there are two types, i.e. BRANE1 and BRANE2.
To satisfy the energy conditions, in the BRANE1, we find the form of
density is expressed as $\rho=3m^2(H^2+2H/l)-\sigma$ and the brane
tension $\sigma$ is positive; while for the BRANE2, the relation of
density is described as $\rho=3m^2(H^2-2H/l)-\sigma$ but it is not
known whether the brane tension is positive. Meanwhile, we get many
inequalities and equations to limit these brane models, and these
relations will be verified by measurement of cosmic fundamental
constants and parameters in the future. At last, we assume arbitrary
matter and use four known energy conditions to study the matter on
the brane in DGP model. If only normal matter is on the brane, we
obtain $-3\leq{q}\leq0$ when $1/r_c\leq{H}\leq5/(4r_c)$ and can not
exclude the accelerating universe. However, If there is nonnormal
matter with $w<-1/3$ on the brane, the universe is accelerated.

\section*{Acknowledgments}
This work was supported by NSF (10573003), NSF (10647110), NSF
(10703001), NBRP (2003CB716300) of P. R. China and DUT 893321.

\end{document}